Original Article

# Towards inferring reactor operations from high-level waste

Benjamin Jung [*], Antonio Figueroa, Malte Göttsche

*Nuclear Verification and Disarmament, RWTH Aachen University, Schinkelstraße 2, Aachen, 52062, Germany*

ABSTRACT

Nuclear archaeology research provides scientific methods to reconstruct the operating histories of fissile material production facilities to account for past fissile material production. While it has typically focused on analyzing material in permanent reactor structures, spent fuel or high-level waste also hold information about the reactor operation. In this computational study, we explore a Bayesian inference framework for reconstructing the operational history from measurements of isotope ratios from a sample of nuclear waste . We investigate two different inference models. The first model discriminates between three potential reactors of origin (Magnox, PWR, and PHWR) while simultaneously reconstructing the fuel burnup, time since irradiation, initial enrichment, and average power density. The second model reconstructs the fuel burnup and time since irradiation of two batches of waste in a mixed sample. Each of the models is applied to a set of simulated test data, and the performance is evaluated by comparing the highest posterior density regions to the corresponding parameter values of the test dataset. Both models perform well on the simulated test cases, which highlights the potential of the Bayesian inference framework and opens up avenues for further investigation.

## 1. Introduction

Fissile material accounting is a cornerstone of the international non-proliferation regime and will likely play an important role in future nuclear disarmament regimes. Verifying an account of a fissile material stockpile is a challenging task if the past operation of the respective nuclear fuel cycle was concealed from the international community. A detailed reconstruction of the fissile material production history will likely be required to ascertain the fidelity of an account (declaration), which is a task that can profit from the application of nuclear archaeology [1]. As a field of research, nuclear archaeology comprises numerous scientific methods for reconstructing the operating histories of nuclear facilities, primarily nuclear reactors and enrichment facilities.

So far, nuclear archaeology methods for nuclear reactors have focused on analyzing samples from permanent structures in the reactor, e.g., graphite moderator or pressure tubes [2,3], to reconstruct operating information relevant to the past plutonium production. However, high-level waste (HLW), typically generated at reprocessing facilities, may also hold insightful information about the operating history of nuclear reactors, and nuclear archaeology would benefit from methods to extract this information.

Reconstructing the origin and history of irradiated nuclear material also falls within the scope of nuclear forensics. In this context, machine-learning-based frameworks have been proposed to reconstruct information from measurements of isotopic ratios in samples of nuclear material [4,5]. These studies included reactor-type classification and irradiation parameter reconstruction. It has been proposed that extending these methods to deal with mixed samples could be possible [4].

Independently, preliminary, simulation-based studies have shown that it is, in principle, possible to use a Bayesian inference framework to reconstruct fuel irradiation parameters from HLW samples [6,7]. Bayesian inference offers the advantage of considering not only the observed evidence but also additional information in the form of a prior. It is versatile because the likelihood can be adjusted to different application cases, and it facilitates quantifying the uncertainties of the reconstructed parameters. The framework was demonstrated with both a single batch of fuel and a mixture of two batches of spent fuel. While promising, the study remained limited in terms of the number of reconstructed parameters and the number of test cases and was restricted to a single reactor type.

In this study, we aim to further develop the Bayesian inference framework as a tool for inferring irradiation parameters of a sample of nuclear waste (spent fuel or HLW), using measurements of intra-element isotopic ratios as evidence. We consider two different models, one focusing on reactor-type classification and the other focusing on reconstructing parameters of a mixture, and apply each model to a set of simulated test data to evaluate its performance. In the following sections, we explain the Bayesian framework, outline the two different reconstruction models, and analyze and discuss their performance on the synthetic test data.

## 2. Bayesian inference

Bayesian inference is a probabilistic method for fitting a model $f$ to data $y$ to obtain a probability distribution on the parameters $\theta$ of the model [8]. A joint probability distribution is derived from the *prior distribution* $p(\theta)$ of the parameters $\theta$ and the *likelihood* $p(y|\theta)$. Bayes' rule yields the *posterior distribution*

$$p(\theta|y) = \frac{p(\theta)p(y|\theta)}{p(y)} \propto p(\theta)p(y|\theta). \quad (1)$$

The prior $p(\theta)$ incorporates all previously held beliefs about the parameters $\theta$, without taking the data into account. The likelihood $p(y|\theta)$, considered to be a function purely of $\theta$, computes the probability of the data $y$ given parameters $\theta$. Since $p(y)$ does not depend on $\theta$, it is treated as a normalization constant, which yields the *unnormalized posterior distribution*

$$p(\theta|y) \propto p(\theta)p(y|\theta). \quad (2)$$

Applied to the context of this paper, the inference parameters $\theta$ are variables of interest, such as *fuel burnup*, *time since irradiation (TSI)*, *initial enrichment* or *reactor type*. The data $y$ are isotopic ratios measured in the spent fuel or HLW. For a given set $\{y_r\}$ of independent isotopic ratios, the likelihood is modeled as:

$$p(y|\theta) = \prod_r \mathcal{N}\left(f_r(\theta) - y_r, \sigma_r\right), \quad (3)$$

with $\mathcal{N}(\mu, \sigma)$ denoting the normal distribution. Here, $f_r(\theta)$ is a computational model that predicts isotopic ratios given a set of parameter values. The model $f$ depends on the nature of the problem and often incorporates implicit assumptions, such as the number of components in a mixed sample. The standard deviation $\sigma_r$ of each ratio captures all sources of uncertainty, such as measurement uncertainty and model uncertainty. We use $\sigma_r = 0.1 \cdot y_r$ for all isotope ratios because a detailed uncertainty analysis is beyond the scope of the present study.

Since Eq. (1) is usually analytically intractable, numerical methods are commonly used to approximate the (unnormalized) posterior distribution. A popular class of methods are *Markov chain Monte Carlo* algorithms, which construct a sequence (called *Markov chain*) of samples that converges to the target distribution. Each sample in the chain depends only on the previous sample, although the exact prescription for deriving a new sample from the previous one depends on the specific algorithm. As these algorithms include a pseudo-random element, one typically creates multiple chains with different starting values to avoid statistical anomalies and enable convergence checking. After sampling, the sample chains are merged to form the samples of the posterior distribution.

In this study, we use the NUTS algorithm as implemented in PyMC [9], a probabilistic programming library for Python. To improve the performance, each chain starts with a tuning phase, during which parameters (e.g., step size) of the algorithm are optimized. The samples drawn during this phase are discarded afterwards and do not count towards the posterior distribution samples.

### 2.1. Surrogate modeling

In the framework described above, the likelihood uses a computational model to calculate isotopic ratios given reactor parameters, a task that is typically done with Monte Carlo reactor physics codes at a rather large computational cost. Additionally, MCMC algorithms evaluate the likelihood very frequently, which leads to an immense computational cost if Monte Carlo neutron transport models are used. Furthermore, the samples are drawn sequentially, as each point in a chain depends on the previous one, which limits the potential for parallelization and thus significantly increases the runtime along with the computational cost.

**Table 1**
Isotopic ratios chosen by the selection algorithm. These ratios were used as evidence for the Bayesian inference test cases.

| Cd | Gd | Nb | Pu | Sm | Sn |
|---|---|---|---|---|---|
| $Cd-112$ | $Gd-154$ | $Nb-94$ [a] | $Pu-242$ | $Sm-151$ | $Sn-126$ |
| $Cd-110$ | $Gd-155$ | $Nb-93m$ | $Pu-241$ | $Sm-154$ | $Sn-117$ |
| $Cd-110$ | $Gd-155$ | $Nb-93$ | $Pu-240$ | $Sm-147$ | $Sn-117$ |
| $Cd-116$ | $Gd-156$ | $Nb-94$ | $Pu-239$ | $Sm-152$ | $Sn-122$ |
| $Cd-110$ | $Gd-158$ | $Nb-93$ [a] | $Pu-238$ | $Sm-147$ | $Sn-122$ |
| – | $Gd-158$ | $Nb-93m$ | $Pu-241$ | $Sm-148$ | $Sn-120$ |
| – | $Gd-160$ | – | – | – | – |
| – | $Gd-157$ | – | – | – | – |

| Ba | Bi | Nd | Pd | Se | Te |
|---|---|---|---|---|---|
| $Ba-132$ [a] | $Bi-213$ [a] | $Nd-146$ | $Pd-107$ | $Se-79$ | $Te-126$ |
| $Ba-137m$ | $Bi-211$ | $Nd-143$ | $Pd-106$ | $Se-82$ | $Te-122$ |
| $Ba-136$ | – | $Nd-144$ | $Pd-104$ | $Se-79$ | $Te-124$ |
| $Ba-134$ | | $Nd-148$ | $Pd-108$ | $Se-77$ | $Te-122$ |

| Dy | Er | I | Pb | Po | Th |
|---|---|---|---|---|---|
| $Dy-161$ | $Er-170$ | $I-127$ | $Pb-209$ [a] | $Po-211$ [a] | $Th-229$ |
| $Dy-164$ | $Er-168$ | $I-129$ | $Pb-214$ | $Po-215$ | $Th-228$ |

[a] These ratios were not used in the classification model due to issues with the surrogate model training.

To address these computational limitations, we use surrogate modeling to replace the Monte Carlo simulation of the reactor physics with a machine learning model that predicts the desired isotope ratios at only a fraction of the computational cost. In previous work, we have shown that Gaussian process regression (GPR) is a suitable method for training such surrogate models [10]. A particular advantage is the relatively low number ($\lesssim 500$) of training data necessary to achieve good model predictions. The training data, i.e. reactor parameter and nuclide density pairs, are generated with Monte Carlo reactor physics calculations using state-of-the-art software packages (SERPENT 2 [11], OpenMC [12] or ONIX [13]).

### 2.2. Isotopic ratio selection

Nuclear waste contains many different elements and isotopes, providing a potential basis for numerous isotope ratios, many of which are not necessarily sensitive to the inference parameters. To obtain meaningful results with Bayesian inference, it is prudent to select a set of suitable isotope ratios. In general, there are two ways of doing this. On the one hand, a careful study of the physical properties governing the time evolution of isotopic ratios can reveal a sensitivity to specific inference parameters. On the other hand, algorithms that compare quantifiable metrics for different sets of isotopic ratios can select the best set without analyzing the physical properties. While both options have their merits, we chose an automated approach in this work to facilitate selecting a larger set of ratios.

In our approach, the selection metric is the maximum of the relative standard deviations $\{\hat{\sigma}_{rel,i}\}$ of the marginal distributions of each parameter $\theta_i$ in a given likelihood $p(y|\theta)$. Each marginal distribution $p(\theta_i)$ is approximated by evaluating the likelihood on a grid of values of $\theta$ and summing over each dimension of the grid except $i$. From each marginal distribution we compute $\hat{\sigma}_{rel,i} = \hat{\sigma}_i/\hat{\mu}_i$ and then calculate the selection metric $\hat{\sigma}_{rel} = \max(\{\hat{\sigma}_{rel,i}\})$. This metric is computed with likelihoods of pairs of ratios, for all possible pair-wise combinations of a set of pre-selected isotopic ratios. The ratio pair with the lowest $\hat{\sigma}_{rel}$ is selected as the best pair. Given that the likelihood is seen as a function of $\theta$ with fixed $y$, we select the best ratio pair for each point $y_{test}$ on a grid of test points that covers the parameter space. The union of these isotope ratio pairs forms the final set of isotopic ratios.

This approach is limited by the curse of dimensionality because grid sampling scales exponentially with the number of dimensions (here: inference parameters). Therefore, we limit the explored parameter space to two inference parameters: burnup and TSI. Table 1 shows the selected isotopic ratios, which we use in this study. Although these isotopic ratios are not strictly independent, we nonetheless treat them as independent in Eq. (3).

One must note that we have only considered a pre-selected set of isotopic ratios where 90% of values from a set of 4000 simulations lie within the interval [1/20, 20], as the precision of measuring small ratios is typically related to counting statistics [14], and we have assumed that these isotopes can be measured.

## 3. Testing inference models

In the absence of actual measurement data from nuclear waste, we use synthetic test data generated by evaluating the Gaussian process (GP) surrogate models on a given set of parameter values, referred to as ground truth. This procedure allows us to evaluate the performance of the subsequent inference in a systematic manner.

We explore two different likelihood models that, although not as complex as a realistic application scenario, represent a significant increase in complexity compared to our previous work. These models illustrate that the general Bayesian approach can be adapted to different use cases that require different model assumptions. The first model reconstructs the source reactor type of a sample of spent fuel or reprocessing waste, as well as some irradiation parameters. This model is referred to as the *classification model*. The second model reconstructs the irradiation parameters of a sample that is a mixture of high-level reprocessing waste from two batches of fuel from the same reactor. This model is referred to as the *mixture model*.

The classification model considers four irradiation parameters: *fuel burnup*, *time since irradiation* (TSI), *initial enrichment* and (average) *power density*. The model considers three different reactor types: a graphite-moderated *Magnox*-type reactor, a pressurized light-water reactor (PWR) and a pressurized heavy-water reactor (PHWR). The Magnox type reactor is modeled after the 5 MWe reactor in North Korea. The PWR is modeled after the German Gemeinschaftskernkraftwerk Neckarwestheim II (GKN-II) and the PHWR is a CANDU-6 model. Training data for the surrogate models was generated for each reactor using the simulation tools `OpenMC` and `ONIX` and the ENDF/B-VIII.0 [15] nuclear data library.

In terms of the inference framework, the model function $f_r$ of one isotope ratio $R_r = \frac{N_i}{N_j}$ (see Eq. (3)) is formulated as follows:

$$f_r(\alpha, \theta) = \frac{\sum_m \alpha_m \cdot F_{i,m}(\theta_m)}{\sum_m \alpha_m \cdot F_{j,m}(\theta_m)}, \quad (4)$$

with $\alpha_m \epsilon \{0, 1\}$ and $m \epsilon \{$Magnox, PWR, PHWR$\}$. The $F_{\{i,j\},m}$ are GP surrogate models that compute nuclide densities for each respective reactor type, given a vector $\theta_m$ of irradiation parameters. The variable $\alpha_m$ encodes the reactor type. It is 0 or 1 depending on which reactor label is sampled, i.e., if PHWR is sampled then $\alpha_{PHWR}$ is 1 and the others are 0. The irradiation parameters are sampled independently for each reactor, as different reactor types have different ranges of possible values. Table 2 shows the limits of the uniform prior distributions of each inference parameter. These ranges reflect typical values for each reactor, with some exceptions. The burnup range of the PWR model is adjusted to the typical values of CANDU and Magnox reactors for the sake of comparability between the three models. The lower limit of the TSI is set to 1000 d given that spent fuel requires a significant time to cool before it can be handled further (e.g. to analyze a sample). The initial enrichment level is varied uniformly so as to cover a broad range of possibilities, although this parameter could be constrained by other sources of information. In practice, these parameter ranges as well as the type of distribution would be modified to reflect the state of knowledge about the sample prior to the analysis.

For this model, the simulated test dataset has 200 test points per "true" reactor type, i.e., a total of 600 test points. For each test point, the ground truth parameter values are pseudo-randomly sampled according to the distributions in Table 2, so as to cover the parameter space evenly.

The mixture model considers a sample composed of waste from two batches of fuel irradiated in a PWR reactor. This model is intended to explore the potential of the Bayesian approach to extract information from samples of mixed waste, although realistic application scenarios would need to consider mixtures of more than two batches.

**Table 2**
Prior distributions of the classification scenario. The reactor type is sampled with a categorical distribution that assigns equal probability to each label. Otherwise, the ranges denote a uniform distribution with $p(x) = \frac{1}{b-a}$ for $x \epsilon [a,b]$ and 0 elsewhere.

|  | Magnox | PHWR | PWR |
| --- | --- | --- | --- |
| Reactor type | $p = 1/3$ | $p = 1/3$ | $p = 1/3$ |
| Burnup [MWd/kg] | 0.1–8 | 0.1–8 | 0.1–8 |
| TSI [d] | 1000–10000 | 1000–10000 | 1000–10000 |
| Power density [kW/l] | 0.01–1 | 1–20 | 20–160 |
| Initial enrichment [%at] | 0.72–1.5 | 0.72–1.5 | 1–5 |

**Table 3**
Limits of the prior distributions used in the mixture scenario. The ranges denote the lower and upper limits of a uniform distribution.

| Parameter | Prior limits |
| --- | --- |
| Burnup 1 [MWd/kg] | 0.5–50 |
| TSI 1 [y] | 0.1–60 |
| Burnup 2 [MWd/kg] | 0.5–50 |
| TSI 2 [y] | 0.1–60 |
| Mixing factor | 0.1–15 |

The inference parameters $\theta_1$ and $\theta_2$ are *burnup* and *time since irradiation* (TSI) of batch 1 and batch 2, respectively, and the model function is:

$$f_r(\alpha, \theta_1, \theta_2) = \frac{\alpha \cdot F_i(\theta_1) + F_i(\theta_2)}{\alpha \cdot F_j(\theta_1) + F_j(\theta_2)}. \quad (5)$$

The parameter $\alpha = n_1/n_2$ is the mixing ratio of the amounts of material of each batch. For example, $\alpha = 0.5$ implies a mixing ratio of 1:2, meaning there is twice the amount of waste from batch 2 in the sample as there is from batch 1.

The training data for the surrogate models $F$ was generated with an infinite lattice model of the Kernkraftwerk Obrigheim implemented in SERPENT 2, using the ENDF/B-VII.1 [16] nuclear data library. The limits of the uniform prior distributions are given in Table 3. They reflect the potential parameter space of a typical PWR used for electricity production.

For the mixture model, we create test cases systematically, in order to study specific combinations of parameters, e.g., a mixture of a "low-burnup" and a "high-burnup" batch with varying mixing factors. These combinations of "true" parameter values are listed in Table 4.

## 4. Results

The performance of the two inference models is assessed in terms of whether the ground truth of a test point was successfully inferred. To determine whether the inference was successful, we compute *highest density regions* (HDRs) of the marginal posterior distribution of each parameter. If the known ground truth of a parameter lies within the HDR, the parameter is considered to have been reconstructed successfully. The HDR is the smallest interval that contains a given proportion (here: 95%) of the probability mass [8]. If the posterior distribution has several separate regions with a high probability density, the HDR is divided into several intervals that together contain at least 95% of the probability mass. In such cases, the reconstruction is considered successful if the ground truth lies in one of the intervals.

### 4.1. Classification model

Fig. 1 shows the results of applying the classification model to a single test point. The bar chart shows the count number of each label in the reactor type samples. In the example, the predicted reactor type is "Magnox", as the number of "Magnox" samples is highest. The "PWR" label was already discarded during the tuning phase of the

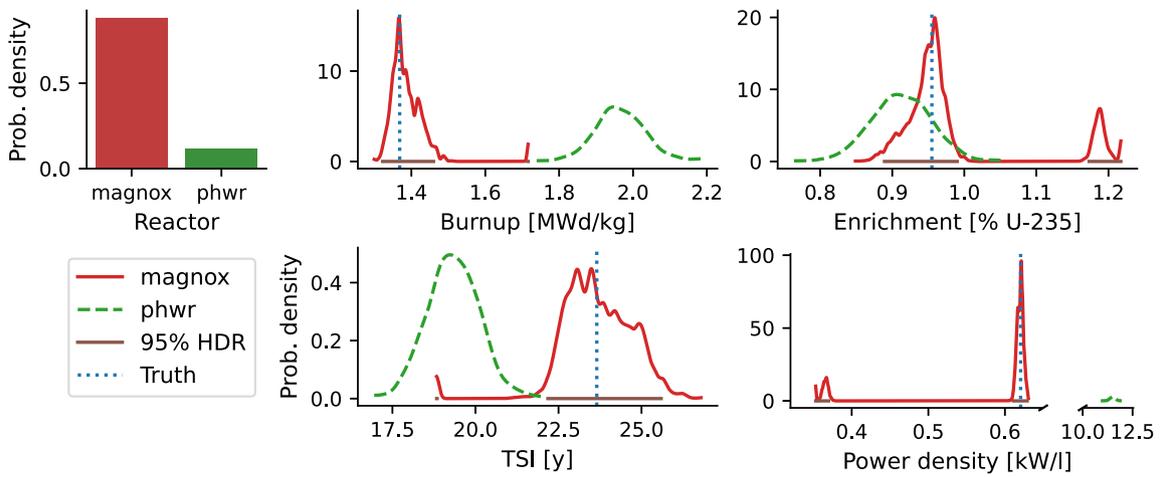

**Fig. 1.** Posterior distributions of a successful classification test run. The bar graph shows the sampled reactor labels and the four graphs below show the marginal posterior distributions of the corresponding parameters. The predicted reactor type is Magnox and horizontal lines indicate the HDR of the posteriors of the Magnox parameters. The true reactor type is Magnox and the true parameter values are indicated by the dotted vertical lines in each graph.

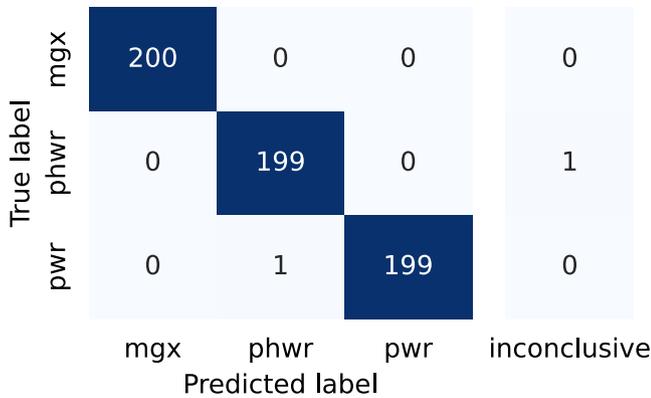

**Fig. 2.** Confusion matrix of the classification model evaluated on the test dataset.

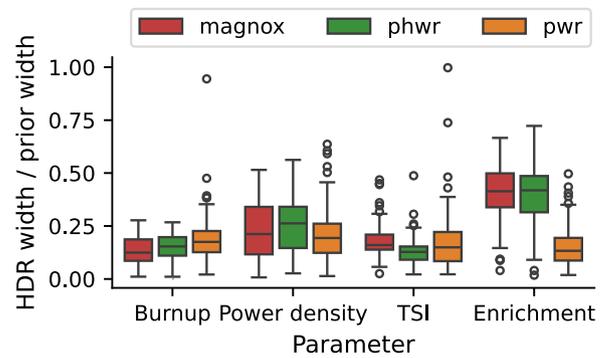

**Fig. 3.** Distribution of cumulative HDR widths relative to the prior widths of the classification model. The cumulative HDR width is the sum of widths of the HDRs of each posterior, and the prior width is the distance between the lower and upper limits of the respective uniform distribution. Includes all test points except the inconclusive test point.

Bayesian inference and does not feature at all. The ground truth lies inside the HDR of all four "Magnox" parameter posteriors, making this an example where the inference works well. One can observe split HDRs in the enrichment and the power density parameters that indicate that a lower power density and higher enrichment are also a potential solution.

Fig. 2 shows the confusion matrix of the reactor type classification. Overall, the misclassification rate is very low. There is only one case of 'inconclusive' classification, in which the highest number of counts of the reactor type samples is shared by two labels. There is also one case of misclassification, in which a 'pwr' sample is mistakenly identified as a 'phwr' sample.

Table 5 tallies the number of successful reconstructions per reactor type and per parameter. Additionally, the table tallies the test points into categories *complete*, *partial* and *none*, meaning either all, some or no parameters were reconstructed successfully. In general, the success rate is high (above 90%). There is no significant difference between the burnup, TSI and enrichment parameters, although the power density parameter has a slightly lower success rate. The PHWR reactor type has the highest rate of complete successes, although the difference to the other two reactor types is small.

In addition to the binary *true/false* success rate, we analyze the widths of the HDRs relative to the widths of the uniform prior distributions. This quantity indicates if the knowledge of a parameter is more precise after (posterior) the inference compared to before (prior). A smaller relative width indicates a greater knowledge gain. Since many test points include posteriors with split HDRs, we define the cumulative HDR width as the sum of the widths of all intervals in a split HDR. Fig. 3 shows boxplots of the cumulative HDR widths relative to the prior widths. Except for a few outliers, all four parameters have relative widths significantly lower than one, indicating a significant gain in information. The spread of relative widths does not vary much across reactor types, except for the enrichment parameter. Here, the PWR test points show much lower relative HDR widths than PWR and Magnox test points. This feature is due to the larger prior width of the PWR type reactor and disappears when comparing the absolute widths.

**Table 4**
Ground truth scenarios of the test dataset for the mixture inference. Each scenario is labeled according to whether the true values are high or low with respect to the range of possible parameters. Each scenario is investigated with a range of different mixing factors: 0.1, 0.2, 0.3, 0.4, 0.5, 0.6, 0.7, 0.8, 0.9, 1, 2, 3, 4, 5, 6, 7, 8, 9, 10, which are applied to the batches denoted with 1.

|        | Burnup 1 [MWd/kg] | TSI 1 [y] | Burnup 2 [MWd/kg] | TSI 2 [y] |
|--------|-------------------|-----------|-------------------|-----------|
| Ll–Ll[a] | 1.5 | 6  | 3   | 3  |
| Lh–Ll  | 3    | 20 | 1.5 | 6  |
| Lh–Lh  | 1.5  | 40 | 3   | 30 |
| Hl–Hl  | 40   | 10 | 45  | 3  |
| Hh–Hl  | 40   | 30 | 45  | 6  |
| Hh–Hh  | 40   | 40 | 45  | 30 |
| Hl–Ll  | 40   | 12 | 1.5 | 6  |
| Hl–Lh  | 40   | 6  | 1.5 | 30 |
| Hh–Ll  | 40   | 30 | 1.5 | 6  |
| Hh–Lh  | 40   | 30 | 1.5 | 25 |

[a] Upper case letters indicate burnup and lower case letters indicate cooling time. H/h stands for a high value and L/l stands for a low value, respective to the range of possible values.

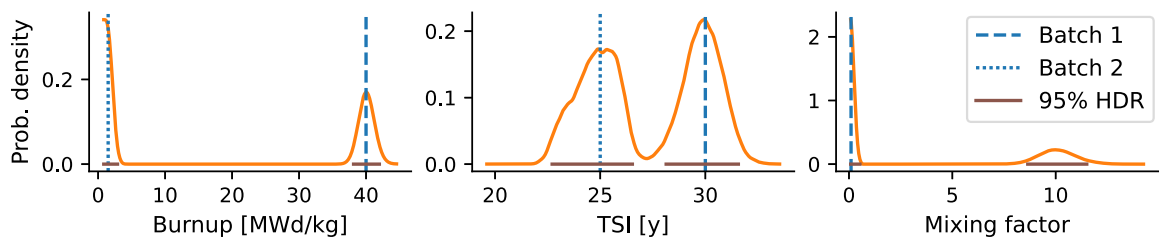

**Fig. 4.** Posterior distributions of the successful mixture test run `Hh-Ll` with $\alpha = 0.1$. The vertical lines indicate the ground truth.

In summary, the classification model performs well on the simulated test data, being able to discriminate between three different reactor models and simultaneously infer irradiation parameters.

*4.2. Mixture model*

The analysis of inference data of the mixture model differs slightly from the classification model. The model function $f$ (see Eq. (5)) treats the sample $(\theta_1, \theta_2, \alpha)$ as equivalent to $(\theta_2, \theta_1, 1/\alpha)$. Therefore, switched parameter values combined with an inverted mixing factor will be treated equally by the sampling algorithm, and the posteriors associated with "batch 1" may contain numerous samples belonging to "batch 2" and vice versa. To determine HDRs, the posterior samples are first merged by parameter type, i.e., the samples of the two burnup parameters are combined into one set and the TSI samples are combined as well, and the HDRs are computed on the joined sets of samples.

Fig. 4 shows an example of such results. The HDRs overlap with the true values, indicating a successful reconstruction. The two intervals of the mixing factor indicate two solutions, around 10 and 0.1 respectively, both of which correspond to the same mixing ratio, i.e. 1:10 or 10:1. One must note that it is also possible to obtain a single HDR for a parameter type. In that case, the reconstruction would be deemed successful if, e.g., both true burnup values lie within this interval.

By calculating HDRs of the marginal distributions independently for each parameter, information about which intervals belong together is lost. That is, one can neither say whether the lower burnup HDR belongs to the lower or the higher TSI HDR, nor can one determine whether the lower or the higher burnup batch has the higher mixing fraction. A potential solution is proposed in the next section.

Fig. 5 shows the inference results for the test points from Table 4 in a color coded matrix. One can observe that the success of the reconstruction depends on the true values of burnup and TSI. Test points with high burnup are reconstructed successfully more often than test points with low burnup. Test points with mixed high and low burnup are generally reconstructed successfully, with some exceptions when the mixing fraction of the low burnup batch is particularly low (see the right upper corner in Fig. 5). The test points with two low burnup batches are mostly not reconstructed successfully. In all of the test points denoted with two "L", 99% of the posterior probability mass of the burnup posterior lies below 15 MWd/kg. Such results can be interpreted to indicate a low burnup, even if a more precise reconstruction of the parameter value does not work.

In summary, the success of the Bayesian inference with the mixture model depends on the nature of the ground truth, particularly on whether the burnup values are high or low, and on the mixing fraction.

**5. Discussion**

This study improves upon previous work on this topic in two ways. First, the "single-batch" model is successfully extended from inferring burnup and TSI of a sample from a known reactor to inferring burnup, TSI, initial enrichment and power density; and discriminating between three different reactor types. Second, we evaluate the performance of the mixture model on a larger set of test cases. The results highlight the potential of using a Bayesian framework to infer information on the origin and the irradiation history of samples of nuclear waste.

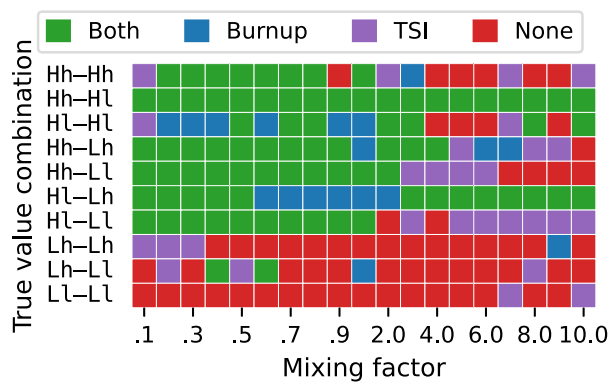

**Fig. 5.** Success matrix of the mixture inference model. The *y*-axis shows the relationship of the true parameter values. Upper case letters indicate burnup and lower case letters indicate cooling time. H/h stands for a high value and L/l stands for a low value, respective to the range of possible values. The colors indicate whether both burnup and TSI, only burnup, only TSI or neither were reconstructed successfully.

**Table 5**
Successful parameter reconstruction in the classification scenarios sorted by true reactor type. 200 test points are considered per reactor type.

|  | Magnox | PHWR | PWR | Total |
|---|---|---|---|---|
| Burnup | 197 | 195 | 194 | 586 |
| Power density | 191 | 195 | 192 | 578 |
| TSI | 195 | 197 | 194 | 586 |
| Enrichment | 199 | 198 | 189 | 586 |
| Complete | 191 | 195 | 187 | 573 |
| Partial | 8 | 3 | 11[a] | 22 |
| None | 1 | 2[b] | 2 | 5 |

[a] The PWR tally includes the misclassified test point in the partial category because burnup and TSI are reconstructed successfully.
[b] The inconclusive result is counted as an unsuccessful reconstruction here.

However, the inference models in this study do not reflect the complexity of the problem that needs to be expected in a realistic application of the framework. Rather, this study is an exploratory analysis to gauge the potential of the framework. To put the results into perspective, there are several factors to address.

In this study, we have used a selection algorithm to choose a set of isotopic ratios, instead of considering the physical attributes of various nuclides. This new selection process lead to a larger set of isotopic ratios being used than in other studies (see [4,5]). Given the successful inference results obtained on the test datasets, especially with the classification model, this approach shows promise as a methodology for selecting suitable isotopic ratios in an automated manner.

The present study treats uncertainties in a simplified manner, ignoring several challenges associated with measuring isotopic ratios. In the likelihood $p(y|\theta)$, both model and measurement uncertainty of an isotopic ratio are accounted for by the standard deviation $\sigma_r$, which we have assumed to be 10% of the ratio value for all ratios. In practice, one would expect different uncertainties for different ratios because the challenges of measuring isotopic ratios with mass spectroscopy, such as isobaric interferences, apply to some isotopes but not all. Similarly, the nuclear data libraries that determine the accuracy of the model predictions are available with varying degrees of precision for different nuclides. Therefore, one should not assume that the results obtained in this simulation-based study with the present list of isotopic ratios will translate directly into practice.

While a detailed study of uncertainty is beyond the scope of this paper, it is worth noting that $\sigma_r$ in Eq. (3) can be determined individually for each measurement. Thus, it is possible to account for measurement and model uncertainty. It is to be expected that different values of $\sigma_r$ will alter the inference results, but we have not systematically investigated this effect.

The inference framework described in this work is not dependent on a specific list of isotopic ratios. The selection process can be repeated and adapted to the circumstances of different application scenarios. Issues of measurability and uncertainty can be addressed either by applying stricter pre-selection criteria or by using ratio-specific $\sigma_r$ in the likelihoods evaluated by the selection algorithm. Future research should investigate the effect of measurement and model uncertainty more rigorously.

Furthermore, the selection algorithm only considered the parameters burnup and TSI, a limitation due to the computational cost of the algorithm. An algorithm that overcomes these issues could consider other parameters and lead to better results. Future studies should seek to further optimize the isotopic ratio selection and could compare different selection methods.

This study has used uniform prior distributions to describe a general lack of knowledge about the parameters. Since priors play an important role in Bayesian analysis, they would be chosen carefully for each different application of the framework. For example, the enrichment level typically depends on the reactor design and could be narrowed down to several options by investigating other sources of information. Then, a non-uniform prior could be used to describe this information, e.g., a series of overlay normal distributions with peaks centered on the possible enrichment levels.

We have attempted to break down the results of each inference into quantities that can be summarized as an overview of the general performance of a likelihood model. The results of Bayesian inference are posterior distributions, which in many cases cannot be easily parameterized and are therefore difficult to assess systematically on a large scale with algorithmic means. In an application context, an analyst could use many different tools to extract information from the posterior, as well as make adjustments to the modeling and sampling algorithm to obtain a nuanced result. For example, the results of an inference with our mixture model could be further analyzed with clustering algorithms, e.g. k-means or Gaussian mixture, to determine the correlation between the reconstructed burnup and cooling time intervals.

## 6. Conclusion

In this exploratory analysis, we have tested the Bayesian inference framework with two different likelihood models, demonstrating the potential and versatility of the Bayesian approach. We have shown that, in principle, it is possible to discriminate between three potential source reactors (Magnox, PWR, and PHWR) and simultaneously infer up to four parameters (burnup, TSI, enrichment and power density). It is also possible to differentiate the irradiation parameters of components of a mixture of radioactive waste from two different irradiation campaigns in a reactor. However, the framework requires further testing and optimization, as well as a rigorous uncertainty analysis before it can fulfill its potential as a versatile tool for reconstructing parameters of interest from samples of nuclear waste. Given that both nuclear archaeology and nuclear forensics benefit from this capability, the methodology described in this work merits further development and the present study provides a basis for future research.

**CRediT authorship contribution statement**

**Benjamin Jung:** Conceptualization, Methodology, Software, Data curation, Investigation, Visualization, Writing – original draft. **Antonio Figueroa:** Conceptualization, Methodology, Software, Data curation, Investigation, Writing – review & editing. **Malte Göttsche:** Conceptualization, Writing – review & editing, Supervision, Funding acquisition.

**Declaration of competing interest**

The authors declare that they have no known competing financial interests or personal relationships that could have appeared to influence the work reported in this paper.

**Acknowledgments**

The authors gratefully acknowledge the funding provided by a Freigeist Fellowship grant of the VolkswagenStiftung.

The authors also gratefully acknowledge the computing time provided to them at the NHR Center NHR4CES at RWTH Aachen University (project number p0020230). This is funded by the Federal Ministry of Education and Research, and the state governments participating on the basis of the resolutions of the GWK for national high performance computing at universities (www.nhr-verein.de/unsere-partner).